\documentclass[amsmath,amssymb,preprint,showpacs]{revtex4}

\usepackage{graphicx}
\usepackage[usenames]{color}

\begin{document}

\title{Pressure dependence of the single particle excitation in the charge-density-wave CeTe$_3$ system}

\author{M. Lavagnini$^{1}$, A. Sacchetti$^{1}$, C. Marini$^{2,3}$, M. Valentini$^{2}$, R. Sopracase$^{2}$, A. Perucchi$^{2,4}$, P. Postorino$^2$, S.
Lupi$^2$, J.-H. Chu$^{5}$, I.R. Fisher$^{5}$, and L. Degiorgi$^{1}$}

\affiliation{$^{1}$Laboratorium f\"ur Festk\"orperphysik,
ETH-Z\"urich, CH-8093 Z\"urich, Switzerland. \\
$^{2}$CNR-INFM-Coherentia and Dipartimento di Fisica Universit\`a ``La
Sapienza'', P.le A. Moro 5, I-00185 Rome, Italy.\\
$^{3}$Dipartimento di Fisica ÒE. AmaldiÓ, Universit\`a degli Studi Roma Tre, via
della Vasca Navale 84, 00146 Roma, Italy and ÒUnit\`a CNISM Roma1Ó,
Universit\`a ``La
Sapienza'', P.le A. Moro 2, I-00185 Roma, Italy.\\
$^{4}$CNR-INFM-Coherentia and Sincrotrone Trieste S.C.p.A., S.S. 14 km 163.5, in Area Science Park, 34012 Basovizza, Trieste, Italy.\\
$^{5}$Geballe Laboratory for Advanced Materials and
Department of Applied Physics, Stanford University, Stanford,
CA 94305-4045, U.S.A.}

\date{\today}

\begin{abstract}
We present new data on the pressure dependence at 300 K of the optical reflectivity of CeTe$_3$, which undergoes a charge-density-wave (CDW) phase transition well above room temperature. The collected data cover an unprecedented broad spectral range from the infrared up to the ultraviolet, which allows a robust determination of the gap as well as of the fraction of the Fermi surface affected by the formation of the CDW condensate. Upon compressing the lattice there is a progressive closing of the gap inducing a transfer of spectral weight from the gap feature into the Drude component. At frequencies above the CDW gap we also identify a power-law behavior, consistent with findings along the $R$Te$_3$ series (i.e., chemical pressure) and suggestive of a Tomonaga-Luttinger liquid scenario at high energy scales. This newest set of data is placed in the context of our previous investigations of this class of materials and allows us to revisit important concepts for the physics of CDW state in layered-like two-dimensional systems.
\end{abstract}



\maketitle

\section{Introduction}
In a strictly one-dimensional (1D) interacting electron system, the Fermi-liquid (FL) state is replaced by a state where interactions play a crucial role, and which is generally referred to as a Tomonoga-Luttinger liquid (TLL) \cite{Luttinger,Tomonaga}. According to the predictions of the TLL theoretical framework \cite{Schulz}, the 1D state is characterized by features such as spin-charge separation and the breakdown of the quasi-particle concepts. The non-FL nature of the TLL is also manifested by the non-universal decay of the various correlation functions. This originates from the power-law behavior $\rho(\omega)\sim\omega^{\alpha}$ ($\omega=E_F-E$) of the density of states (DOS) close to the Fermi level $E_F$, as immediate consequence of the absence of the discontinuity at $k_F$ in the momentum distribution function. The exponent ${\alpha}$ reflects the nature and strength of the interaction. It is also worthwhile to emphasize that the TLL, which in principle describes so-called gapless 1D fermion systems, may be unstable towards the formation of a spin or a charge gap \cite{Voit,Giamarchi}. Spin gaps are usually considered in microscopic 1D models including electron-phonon coupling, and are relevant to the description of the normal state of superconductors and Peierls insulators \cite{Voit}. Charge gaps, on the other hand, are a more typical consequence of electronic correlations \cite{Schulz}.

Recently, a great deal of interest has been devoted to the possible breakdown of the FL framework in quasi one-dimensional materials. Of paramount importance in this regard, are the linear-chain organic Bechgaard salts, which were indeed intensively investigated and on which the very first optical signature of a TLL behavior was recognized \cite{Vescoli,VescoliEPJ}. Upon compressing the lattice by chemical substitution of the organic molecules and/or counter-ions, the Bechgaard salts display a dimensionality-driven crossover from a 1D Mott-insulator to an incipient 2D Fermi-liquid \cite{Vescoli,Jerome}. These findings were later confirmed by optical experiment on the representative (TMTTF$_2$)AsF$_6$ system under externally applied pressure \cite{Pashkin}. This is associated to a confinement-deconfinement crossover induced by the enhancement of the interchain coupling (i.e., the charge transfer integral, $t_{perp}$), which, besides inducing a warping of the Fermi surface (FS), theoretically implies a self-doping process in these materials \cite{Giamarchi}. Interestingly enough, we identified in our optical data at low temperatures on linear-chain organic materials two relevant energy intervals: a low one (i.e., $\omega < E_g$, $E_g$ being the Mott-Hubbard correlation gap) dominated by the Drude term due to the itinerant charge carriers as consequence of the self-doping, and a high one ($\omega > E_g$, $t_{perp}$) for which an asymptotic one-dimensional limit is achieved and a characteristic non-universal power-law behavior in the real part $\sigma_1(\omega)$ of the optical conductivity ($\sigma_1(\omega)\sim \omega^{-\eta}$) was recognized \cite{Vescoli,VescoliEPJ,Jerome}. It is worth warning the reader, that the TLL theory, applied here, emerged $de facto$ from a truly 1D scenario \cite{Luttinger,Tomonaga,Schulz}, since a rigorous theoretical approach, accounting for the dimensionality crossover, is still missing. Thus, caution should be placed in addressing situations approaching a 2D limit.

FL theory has been thoroughly tested on a variety of materials and is usually valid in higher than one dimension. Nevertheless, this notion seems to break down in several notable exceptions, like in several correlated metals and in the copper oxide-based high-temperature superconductors (HTSC) \cite{Proceedings}. Focusing on the 2D layered-like HTSC, this was beautifully shown from the perspective of the optical response by the recent investigation reported in Ref. \onlinecite{Basov}. Indeed, a similar behavior of $\sigma_1(\omega)$ at high energies, as in the quasi-1D organic materials, was revealed. The high frequency tail of the mid-infrared feature in HTSC decays in a power-law like fashion. This leads to the question, to which extent the effective dimensionality of the interacting electron gas affects the electronic properties in those metals. There is then the quest for alternative layered-like systems, not affected by strong correlations and thus allowing a broader and more general perspective on these issues. 

The two-dimensional rare-earth tri-telluride compounds $R$Te$_3$ ($R$=La, Ce, Pr, Nd, Sm, Gd and Dy) are well suited to that purpose \cite{Dimasi}. Their crystal structure is made up of square planar Te sheets and insulating corrugated $R$Te layers which act as charge reservoirs for the Te planes. They exhibit an incommensurate CDW, residing within the Te planes and stable across the available rare-earth series \cite{Dimasi,Ru1,Ru2}. In our first optical investigations on this family of compounds, we established the excitation across the CDW gap and discovered that this gap is progressively reduced upon compressing the lattice either with chemical substitution (i.e., by changing $R$) or with externally applied pressure \cite{Sacchettiprb,Sacchettiprl}. For energies larger than the CDW gap, we also observed power-laws in $\sigma_1(\omega)$ (i.e., $\sigma_1(\omega)\sim \omega^{-\eta}$), again consistent with the typical behavior of a Tomonaga-Luttinger liquid system. This finding anticipates that interactions could play a significant role in shaping the electronic properties of these CDW materials at such high frequency scales. The direct interaction between electrons can directly produce an umklapp scattering, allowing the transfer of two particles from one sheet of the FS to the other and thus transferring $4q$ to the lattice. The advocated wave vector \cite{Brouet,BrouetPRB} for the CDW modulation is indeed giving a value for $4q$ close to a reciprocal lattice vector, so that such umklapp processes are effective \cite{Sacchettiprb}. Nonetheless, it was recognized, that more studies both theoretically and experimentally would of course be useful in order to ascertain the respective roles of the interactions and of the electron-phonon coupling with respect to the CDW formation. As for the latter coupling, we recently addressed the issue of the interplay between electronic and phononic degrees of freedom in $R$Te$_3$ by Raman scattering investigations \cite{LavagniniRaman}. We found that the lattice dynamics is quite of relevance towards the formation of the CDW condensate and primarily affects the low energy scales. 

In this contribution, we will focus again our attention on the electronic excitation spectrum (i.e., the CDW gap) and on the impact of the direct interaction between electrons, which is supposed to govern the high-frequency properties of $R$Te$_3$ and therefore the shape of $\sigma_1(\omega)$ above the CDW gap \cite{Sacchettiprb}. Emphasis will be also placed on the possibility to systematically tune such materials so that a dimensionality as well as a confinement-deconfinement crossover may be induced. The prerequisite for a reliable study of the high-frequency behavior in $\sigma_1(\omega)$ is the capability to achieve the optical response over a very broad spectral range, which extends well above the energy scale of the gap excitation. This was possible so far for the rare-earth $R$Te$_3$ series, where the physical properties can be varied by lattice compression, achieved through chemical substitution. A more direct and cleaner way to study the electronic properties upon lattice compression is the application of external pressure, since the changes in the electrodynamic response can be monitored while the CDW state is continuously suppressed by tuning the interchain coupling and altering the nesting conditions upon decreasing the lattice constants. Our first investigations of the optical response as a function of applied pressure on CeTe$_3$ \cite{Sacchettiprl} allowed us to address yet the pressure dependence of the CDW gap but were covering a too small energy interval to allow addressing the high-frequency decay of $\sigma_1(\omega)$.

Here, we present new optical data collected as a function of pressure in CeTe$_3$ over a broad spectral range going well beyond the CDW gap and extending up to 1.5x10$^4$ cm$^{-1}$. We have two major targets in mind. Besides revisiting the evolution of the CDW gap upon hydrostatically compressing the lattice and its implication on the electronic properties, we want to achieve a robust determination of the frequency dependence of the optical conductivity at high energy scales, looking for the pressure dependence of its possible power-law behavior. This is supposed to supply a fingerprint on the evolution of the intrinsic dimensionality. We will first introduce the experimental technique and present the data. This will be followed by a short description of the analysis procedure and by a discussion of the resulting electrodynamic response, emphasizing particularly a scenario within the TLL framework. Our final goal is the comparison of the high-frequency optical response in $R$Te$_3$ as a function of both chemical and applied pressure.

\section{Experiment and Results}

The CeTe$_3$ single crystal was grown by slow cooling a binary melt, as already described elsewhere \cite{Ru1}. At ambient pressure we first collected optical reflectivity data on a characterized CeTe$_3$ sample over the broad spectral range, extending from the far infrared (FIR) up to the ultraviolet (UV) and obtaining results fully equivalent to our previous investigations \cite{Sacchettiprb}. A small piece (i.e., 50x50 $\mu$m$^2$) of that specimen was cut and placed on the top surface of NaCl, acting as pressure medium in the stainless steel gasket hole of the pressure cell. We made use of the same experimental set-up, already described in Ref. \onlinecite{Sacchettiprl}. A clamp-screw diamond anvil cell (DAC) equipped with high-quality type IIa diamonds (400 $\mu$m culet diameter) was employed for generating high pressures up to 9 GPa. Pressure was measured with the standard ruby-fluorecence technique \cite{Mao}.

We carried out optical reflectivity ($R(\omega)$) measurements as a function of pressure exploiting the high brilliance of the SISSI infrared beam-line at ELETTRA synchrotron in Trieste \cite{SISSI}. We explore the spectral range between 3200 and 1.5x10$^4$ cm$^{-1}$ \cite{comment1}. The investigated frequency interval was covered with a Bruker Michelson interferometer equipped with a CaF$_2$ beamsplitter as well as HgCdTe and Si detector for the energy intervals 3200-11000 and 9000-16000 cm$^{-1}$, respectively.

An important issue when measuring the reflectivity inside the DAC concerns the precise determination of the reference signal, which has important implications on the correct shape and absolute value of the resulting $R(\omega)$. Similarly to our first investigation \cite{Sacchettiprl}, we measured at each pressure the light intensity reflected by the sample $I_S(\omega)$ and by the external face of the diamond window $I_D(\omega)$, thus obtaining the quantity $R_D^S(\omega)=I_S(\omega)/I_D(\omega)$. At the end of the pressure run, we also measured the light intensity reflected by a gold mirror ($I_{\textrm{Au}}(\omega)$) placed between the
diamonds at zero pressure and again $I_D(\omega)$, acting as a reference. One achieves $R_D^{\textrm{Au}}(\omega)=I_{\textrm{Au}}(\omega)/I_D(\omega)$, which is assumed to be pressure independent. This
procedure allows us to finally obtain the sample
reflectivity $R(\omega)=R^S_D(\omega)/R_D^{\textrm{Au}}(\omega)$ at each pressure, which takes into account the variations in the light intensity due to the smooth depletion of the current in the storage ring. As we have shown previously \cite{Sacchettiprl}, there is a difference between the absolute value of the collected $R(\omega)$ data and that of the expected $R(\omega)$ calculated from the complex refractive index at zero pressure \cite{Sacchettiprb} and assuming the sample inside the DAC \cite{Wooten,Dressel,simulation}. Such a difference can be ascribed to diffraction effects induced by the non-perfectly flat shape of the sample. This latter issue was taken into account by defining a smooth (pressure independent) correction function which is then applied to all spectra  \cite{Sacchettiprl}.

\begin{figure}[!tb]
\center
\includegraphics[width=8.2cm]{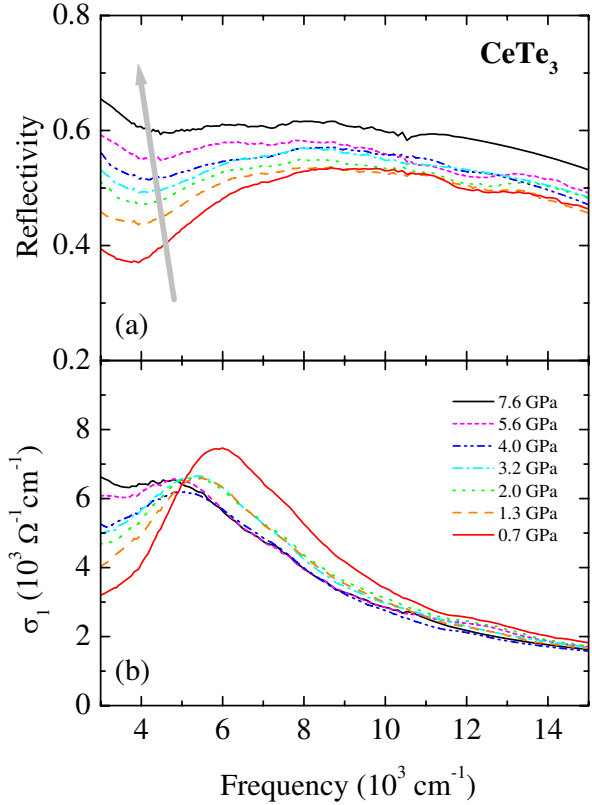}
\caption{(color online) (a) The reflectivity $R(\omega)$ of CeTe$_3$ at 300 K as a function of pressure and (b) the real part $\sigma_1(\omega)$ of the optical conductivity (see text). The arrow in panel (a) indicates the $R(\omega)$ trend upon increasing pressure.}
\label{Refl}
\end{figure}

Figure 1a shows the collected $R(\omega)$ data on CeTe$_3$ while Fig. 1b displays the resulting $\sigma_1(\omega)$ (see below for the detailed procedure) as a function of pressure at 300 K in the spectral range covered by the measurement with the sample inside the DAC. Besides extending to much higher frequencies than in our first investigation \cite{comment1}, the quality of the present data is also very much improved and the annoying interference pattern due to diffused light in our previous spectra \cite{Sacchettiprl} are no longer present. The overall trend agrees, however, with the early findings and establishes once more an analogy between the optical responses upon compressing the lattice either by externally applied pressure \cite{Sacchettiprl} or by chemical substitution \cite{Sacchettiprb}. $R(\omega)$ at low pressures is characterized by the depletion at about 4000 cm$^{-1}$, at the onset of the broad bump which extends into the UV spectral range. The minimum of $R(\omega)$ at about 4000 cm$^{-1}$ progressively fills in under pressure. The broad bump is ascribed to the excitation across the CDW gap into a single particle (SP) state \cite{Sacchettiprb,Sacchettiprl}. Its disappearance with pressure signals the closing of the CDW gap and is accompanied by an overall enhancement of $R(\omega)$, indicating an enhancement of the metallicity of the system. These latter features and behaviors are by now well established and are common experimental facts in the $R$Te$_3$ series. An alternative way to emphasize the trend of $R(\omega)$ in the gap region is to consider the $R(\omega)$ ratio between data at various pressures with respect to those at the highest pressure. Figure 2 shows such a ratio for CeTe$_3$ as well as for previously collected data on NdTe$_3$, further underlying the progressive suppression of the CDW gap upon compressing the lattice. Furthermore, we can appreciate that the depletion of $R(\omega)$ in NdTe$_3$  is, as expected, less pronounced in this spectral range than in CeTe$_3$. This is the direct consequence of the smaller gap in the NdTe$_3$ than in the CeTe$_3$ compound, as achieved through chemical pressure \cite{Sacchettiprb}.

\begin{figure}[!tb]
\center
\includegraphics[width=8.2cm]{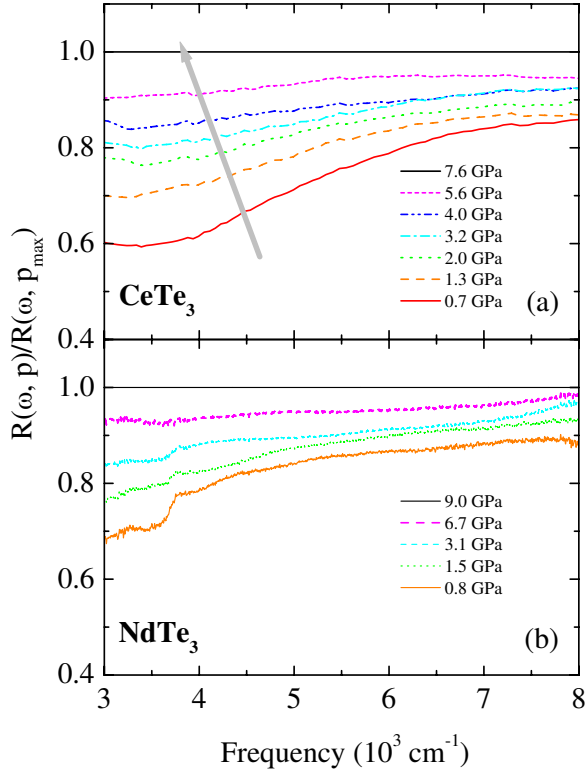}
\caption{(color online) Reflectivity ratio $R(\omega,p)/R(\omega,p_{max})$ at 300 K for CeTe$_3$ (a) and for NdTe$_3$ (b). $p_{max}$ corresponds to the maximal pressure of 7.6 GPa and 9 GPa reached in our experiment for CeTe$_3$ and NdTe$_3$, respectively. The arrow in panel (a) indicates the $R(\omega)$ trend upon increasing pressure.}
\label{Refl}
\end{figure}

\section{Analysis}
Before applying the Kramers-Kronig transformations in order to get the optical conductivity one first needs to extend $R(\omega)$ beyond the measured energy interval. We made use of the same analysis successfully applied in our recent work on LaTe$_2$ \cite{Lavagniniprbpressure}, based on the well established procedure described in Ref. \onlinecite{Pashkin}. For the sake of completeness, we just briefly summarize the essential steps, referring to the quoted literature for more details.

First of all, it is worth recalling that the complete electrodynamic response of CeTe$_3$ at ambient pressure over the whole FIR-UV spectral range can be consistently fitted within the Lorentz-Drude (LD) approach \cite{Wooten,Dressel}; namely, with one Drude term for the metallic contribution, three Lorentz harmonic oscillators (h.o.) for the broad CDW gap feature and two additional h.o.'s for the onset of the electronic interband transitions \cite{Sacchettiprb}. With these fit components it is possible to reproduce $R(\omega)$ under pressure as well, by fitting the data in the measured energy interval only and obviously accounting for the sample inside the DAC \cite{simulation}. The quality of the resulting fit is nevertheless satisfactory despite of the limited spectral range effectively fitted, as exemplified in Fig. 3a for the data taken at 0.7 GPa. This allows us to extrapolate the $R(\omega)$ spectra beyond the experimentally available energy interval (Fig. 3a). Figure 3b displays the optical conductivity calculated within the same Lorentz-Drude approach applied to $R(\omega)$ (Fig. 3a) and shows the related single components of the fit procedure, as well. 

The precise frequency dependence of $\sigma_1(\omega)$ is sensitively governed by subtle changes of the measured $R(\omega)$. In order to reconstruct $\sigma_1(\omega)$ free from any constrains imposed by the Lorentz h.o.'s, we perform reliable Kramers-Kronig (KK) transformation of the extended $R(\omega)$ (Fig. 3a). For the sample-diamond interface the KK relation for the phase $\phi$ of $R(\omega)$ is given by \cite{Plaskett, McDonald}:
\begin{equation}
  \phi(\omega_0)=-\frac{\omega_0}{\pi}P \int_{0}^{+\infty}\frac{\ln R(\omega)}{\omega^2-\omega_0^2}d\omega+\left[\pi-2\arctan\frac{\omega_\beta}{\omega_0}\right],
  \label{eq:KK}
\end{equation}
where $\omega_\beta$ is the position of the reflectivity pole on
the imaginary axis in the complex frequency plane. Analogous to Ref. \onlinecite{Lavagniniprbpressure}, $\omega_\beta$ is chosen so that  $\sigma_1(\omega)$ from KK and from the Lorentz-Drude fit agree \cite{Pashkin}. Table I summarizes $\omega_\beta$ at all investigated pressures.

\begin{figure}[!tb]
\center
\includegraphics[width=8.2cm]{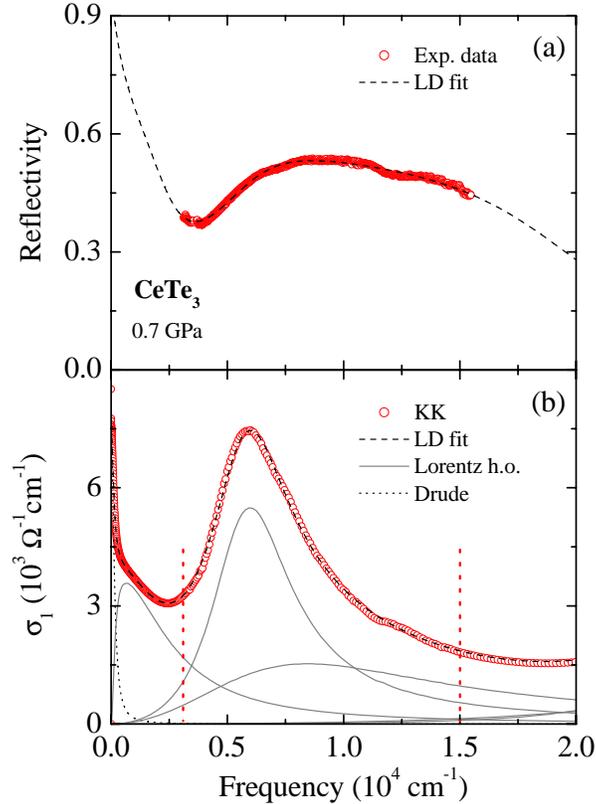}
\caption{(color online) (a) Measured $R(\omega)$ of CeTe$_3$ at 0.7 GPa and its extension based on the Lorentz-Drude (LD) fit (see text). (b) Real part $\sigma_1(\omega)$ of the complex optical conductivity achieved through Kramers-Kronig (KK) transformation of the spectrum in panel (a) and its reproduction within the Lorentz-Drude fit. The fit components are displayed, as well. The dashed vertical lines in panel (b) highlight the spectral range, where the original $R(\omega)$ data were collected.}
\label{Refl}
\end{figure}

Figure 3b compares $\sigma_1(\omega)$ from KK and from the direct Lorentz-Drude fit. The astonishingly good agreement between the two procedures well testifies for the self-consistency of the applied analysis. Similarly to our previous work reported in Ref. \onlinecite{Lavagniniprbpressure}, we took good care to check the impact of the high frequency extrapolations of $R(\omega)$ on the KK results. There is no noticeable effect of the different extrapolations of $R(\omega)$ on $\sigma_1(\omega)$ in the measured spectral range. We are then confident that the data and particularly their analysis within the KK procedure sketched here are robust in the spectral range covered experimentally and can be trusted all the way up to the experimental limit of approximately 15000 cm$^{-1}$.

\begin{table}[!t]
\centering
\begin{tabular*}{\columnwidth}{@{\extracolsep{\fill}}cccccc}
\\\hline\hline
p (GPa) & $\omega_{\beta} (cm^{-1})$ & $\omega_{SP} (cm^{-1})$ & $\omega_{p} (cm^{-1})$ & $\Phi$
& $ \eta$ \\
\hline
0.7 & 31500 & 5933 & 8000 & 0.020& 1.6 \\
1.3 & 31000 & 4933 & 9000 & 0.026 & 1.5 \\
2.0 & 32000 & 4605 & 10000 & 0.031 & 1.4 \\
3.2 & 32000 & 4144 & 10000 & 0.037 & 1.4 \\
4.0 & 32000 & 3682 & 10500 & 0.035 & 1.4 \\
5.6 & 33000 & 2811 & 12000 & 0.042 & 1.3 \\
7.6 & 34000 & 2121 & 12500 & 0.043 & 1.3 \\

\hline\hline
\end{tabular*}
\caption{Pressure dependence of the reflectivity energy pole $\omega_{\beta}$, the single particle peak $\omega_{SP}$, the plasma frequency $\omega_p$, the fraction $\Phi$ of the un-gapped Fermi surface and the power-law exponent $\eta$.} \label{Tab}
\end{table}

\section{Discussion}

Figure 1b presents $\sigma_1(\omega)$, obtained through the KK procedure sketched above (Fig. 3) and in the corresponding spectral range where the $R(\omega)$ spectra (Fig. 1a) were originally collected. We immediately recognized two features: the single particle peak excitation due to the CDW gap and the finite $\sigma_1(\omega)$ at low frequencies which defines the onset of the effective metallic contribution. Upon applying pressure it is evident that spectral weight is removed from the CDW gap excitation and moves into the Drude term at low frequencies. This is coincident with the already quoted filling-in of the $R(\omega)$ depletion at about 4000 cm$^{-1}$ (Fig. 1a). Furthermore, one can qualitatively appreciate the shift of the CDW gap excitation towards low frequencies, indicating its progressive closing upon compressing the lattice. We like to recall that such a behavior of $\sigma_1(\omega)$ is totally consistent with the findings in the chemical pressure experiment \cite{Sacchettiprb} as well as in the related LaTe$_2$ compound under externally applied pressure \cite{Lavagniniprbpressure}. In contrast to our first set of data on CeTe$_3$ under pressure \cite{Sacchettiprl}, the present data extend to high enough frequencies \cite{comment1}, so that the high-frequency tail of the CDW gap absorption can be esteemed over the measured energy interval. The upturn in $\sigma_1(\omega)$ at the high frequency limit of our spectra (i.e, at about 2x10$^4$ cm$^{-1}$, Fig. 3b) signals the onset of the electronic interband transitions.

The wide investigated spectral range allows us a precise determination of the single particle peak excitation $\omega_{SP}$, then ascribed to the CDW gap. The Lorentz-Drude fit gives us access to the various phenomenological components pertinent to the energy interval considered here and ultimately allows to disentangle the (re-)distribution  of spectral weight among them as a function of pressure. Analogous to our previous work \cite{Sacchettiprb,Sacchettiprl,Lavagniniprbpressure} we define the CDW gap energy as the average weighted energy:
\begin{equation}
\omega_{SP}=\frac{\sum_{j=1}^3 \omega_j S_j^2}{\sum_{j=1}^3
S_j^2},
\end{equation}
where $\omega_j$ is the resonance frequency and $S_j^2$ is the strength of the $j$-Lorentz h.o. The sum is over the first three h.o.'s (Fig. 3b). The values of $\omega_{SP}$ are summarized in Table I and fairly agrees with our estimations based on the first data set presented in Ref. \onlinecite{Sacchettiprl}. The decrease of $\omega_{SP}$ upon hydrostatically compressing the lattice is quite evident. Nonetheless, we remark that $\omega_{SP}$ decreases in a slightly slower manner than in Ref. \onlinecite{Sacchettiprl}. This is due to the unavoidable uncertainties, particularly in the determination of the low frequency spectral weight, since only the onset of the metallic contribution can be observed in the experiment with applied pressure. Nevertheless, guided by the trend of $\omega_p$ established in the $R$Te$_3$ series \cite{Sacchettiprb}, we can exploit the onset of the metallic component in $\sigma_1(\omega)$ for a rough estimation of the Drude weight in CeTe$_3$ under pressure. The plasma frequency values related to the Drude term are also reported in Table I.

Instead of treating the single energy scales separately, it is more appropriate and instructive to discuss them from the perspective of sum rule arguments. We can indeed estimate the widely used ratio:
\begin{equation}
\Phi=\frac{\omega_p^2}{(\omega_p^2+\sum_{j=1}^3 S_j^2)}
\end{equation}
between the Drude weight in  the CDW state (i.e., $\sim \omega_p^2$) and the total weight of the hypothetical normal state (i.e., when the weight encountered in the CDW gap ($\sum_{j=1}^3 S_j^2$) merges into the Drude term) \cite{Sacchettiprb,Sacchettiprl,Lavagniniprbpressure}. This latter equation, following well established concepts employed elsewhere \cite{Perucchi}, tells us how much of the FS survives in the CDW ground state and is not gapped by the formation of the CDW condensate. The estimation of $\Phi$ for the $R$Te$_3$ series (i.e., chemical pressure) \cite{Sacchettiprb} relies on the very robust Lorentz-Drude analysis of the electrodynamic response over a very broad spectral range, covered with optical experiments at ambient pressure. We established that as little as 2\% to 7\% of the original FS remains in the CDW state, going across the chemical series from LaTe$_3$ to DyTe$_3$ \cite{Sacchettiprb}. These values are smaller but of the same order of magnitude than those obtained through a de Haas-van Alphen investigation of LaTe$_3$ \cite{RuHvA}. Although the agreement among the two techniques is not perfect, these latter data give some confidence in our procedure based on spectral weight arguments of $\sigma_1(\omega)$ in establishing the un-gapped fraction of the FS. For the optical experiment under applied pressure, caution should be placed on the estimation of $\Phi$, which is obviously affected by the limitations in estimating the Drude weight, as pointed out above (i.e., this is accounted for by the systematically larger error in CeTe$_3$ under pressure than in the $R$Te$_3$ series). Values of $\Phi$ as a function of pressure are summarized in Table I, and are found to be in agreement with those of the chemical pressure experiment. Figure 4 compares the relationship between $\omega_{SP}$ and $\Phi$ upon compressing the lattice, both chemically \cite{Sacchettiprb} and hydrostatically. The overall similar trend of $\Phi$ versus $\omega_{SP}$ in both experiments is pretty obvious. On the one hand, this indicates that the analysis of the spectral weight redistribution in terms of $\Phi$ is also reliable for the applied pressure experiment and on the other hand, this generally emphasizes the increase of the metallicity in $R$Te$_3$ upon suppressing the CDW condensate. Closing the gap releases additional charge carriers in the conducting channel so that the fraction of the un-gapped FS increases.

\begin{figure}[!tb]
\center
\includegraphics[width=9cm]{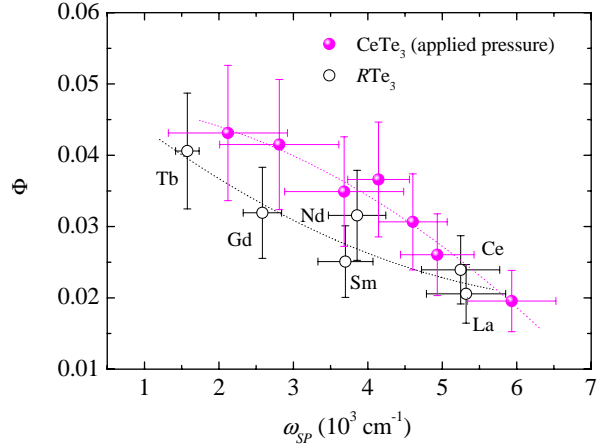}
\caption{(color online) The ratio $\Phi$ (eq. (3)) versus the single particle peak $\omega_{SP}$ (eq. (2)) for the experiment on CeTe$_3$ under externally applied pressure and for the data collected on the $R$Te$_3$ series (i.e., chemical pressure) \cite{Sacchettiprb}.}
\label{Refl}
\end{figure}

As pointed out above, $\sigma_1(\omega)$ is totally unaffected by the extrapolation of $R(\omega)$ above the upper experimental limit at about 1.5x10$^4$ cm$^{-1}$, necessary for the purpose of the KK transformation. This allows a robust assessment of the shape and frequency dependence of $\sigma_1(\omega)$ as a function of pressure at frequencies larger than the CDW gap. Figure 5 highlights $\sigma_1(\omega)$ for the spectral range above the CDW gap, using a bi-logarithmic scale representation. Furthermore, $\sigma_1(\omega)$ has been rescaled by its maximum value and the frequency axis by the frequency $\omega_{max}$ where the maximum of $\sigma_1(\omega)$ occurs. The high energy tail of the CDW gap is well reproduced by a power-law behavior $\sigma_1(\omega)\sim\omega^{-\eta}$
 with exponent $\eta$ between 1.6 and 1.3 (Table I) \cite{comment2,commenteta,exponent}. The overall behavior of $\sigma_1(\omega)$ is very similar to the data on the linear-chain organic Bechgaard salts \cite{Vescoli,VescoliEPJ,Jerome} and agrees with findings in the rare-earth series $R$Te$_3$ \cite{Sacchettiprb} as well as with our most recent data collected on LaTe$_2$ under pressure \cite{Lavagniniprbpressure}. Therefore, we are confident that the power-law behavior is here a reliable experimental fact. 
 
\begin{figure}[!tb]
\center
\includegraphics[width=9cm]{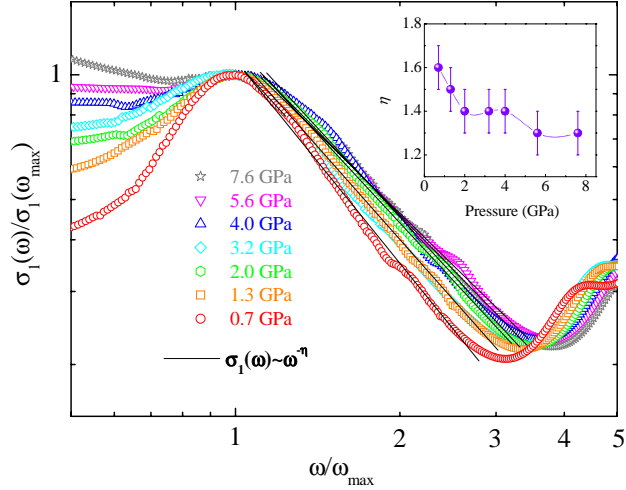}
\caption{(color online) $\sigma_1(\omega)$ of CeTe$_3$ at selected pressures plotted on a bi-logarithmic scale. The y-axis is scaled by the maximum of the mid infrared peak in $\sigma_1$, while the energy axis is scaled by the frequency ($\omega_{max}$) where the maximum in $\sigma_1(\omega)$ occurs. The solid lines are power-law ($\sigma_1(\omega)\sim\omega^{-\eta}$) fits to the
data. The pressure dependence of the exponent $\eta$ is summarized in the inset \cite{commenteta}, as well as in Table I.}
\label{Refl}
\end{figure}

In the case of a 1D material, one would predict different exponents for the optical behavior above the Peierls gap, depending on the hierarchy of the energy scales as well as on the leading interaction involved in the formation of the CDW condensate \cite{Giamarchibook}. As anticipated in the introduction, the observed power-law behavior in $\sigma_1(\omega)$ of the $R$Te$_3$ series would be consistent with predictions based on the Tomonaga-Luttinger liquid scenario, with umklapp scattering processes playing a rather important role \cite{Sacchettiprb,Giamarchi,exponent}. Therefore, the direct electron-electron interaction rather than the electron-phonon one would dominate in this case at such high energy scales above the CDW gap. Identifying the behavior of $\sigma_1(\omega)$ within the TLL scenario further implies that these systems and specifically their electronic properties are shaped by a hidden one dimensionality, despite their 2D layered-like structure \cite{Sacchettiprb}. The issue of the effective dimensionality of the interacting electron gas is quite of relevance here. The advocated FS of $R$Te$_3$ consists of two sheets of open FS's of a quasi 1D material (associated to the $p_x$ and $p_y$ orbitals, respectively). The measured vector for the CDW modulation is very
close to the vector that corresponds to the nesting of the two
sides of this quasi-1D FS. Attributing the one dimensionality to
the fact that $R$Te$_3$ have a nearly perfect nested quasi-1D FS
plays a decisive role, since the charge transfer
integral ($t_{perp}$) along the transverse direction (i.e.,
describing the hopping between the $p_x$ and $p_y$ orbitals) is
not small and is much larger than the temperature of the
measurements. Indeed, $t_{perp}>T$ would normally lead to coherent
transverse hopping, so that FS would have significant warping in
the transverse direction and the material would not be 1D anymore.
The warping of the FS would then loose its relevance only at
$\omega>t_{perp}$. However, this is not the appropriate situation
for $R$Te$_3$, since $t_{perp}= 0.37$ eV \cite{Brouet,BrouetPRB}, while the power-law behavior (Fig. 5) is observed already for frequencies $\omega\sim t_{perp}$. But
if nesting is strong and occurs with a well defined $\vec{q}$
vector, then the system still acts as a 1D system would
essentially do. The 1D character, indicated by the high frequency
power-law behavior of $\sigma_1(\omega)$ (Fig. 5), may then
persist even for $\omega\sim t_{perp}$, provided that one
looks at phenomena involving the nesting wave-vector.

Finally, it is worth pointing out that the exponent $\eta$ progressively decreases upon compressing the lattice (inset in Fig. 5) which would be quite suggestive of a crossover from a weakly interacting towards a non-interacting electron gas system upon reducing the lattice constant \cite{commenteta}. Such a trend in $\eta$ is not uncommon and has been most clearly observed in the linear-chain Bechgaard salts \cite{Vescoli,Pashkin}. Furthermore, it is worth recalling that, apart from the exception of LaTe$_3$, $\eta$ behaves similarly in the chemical pressure experiment across the $R$Te$_3$ series \cite{Sacchettiprb}.  Nevertheless, the trend of $\eta$ in CeTe$_3$ (inset of Fig. 5) is different from what has been observed in LaTe$_2$ under pressure \cite{Lavagniniprbpressure}. In this latter material, $\eta$, while being also quite close to 1, remains basically constant as a function of pressure. We may speculate that pressure on the double layered $R$Te$_3$ is more effective than on the single layered $R$Te$_2$. This may imply a more pronounced (1D to 2D) dimensionality crossover upon applying pressure in the $R$Te$_3$ than in the $R$Te$_2$ series. 

\section{Conclusions}
We have reported a comprehensive infrared optical investigation of CeTe$_3$ under externally applied pressure. We achieved two major results. First, we could revisit the pressure dependence of the CDW gap, confirming its closing upon compressing the lattice. The gap closing is obviously accompanied by the enhancement of the FS fraction accounting for the metallic state. The second major achievement is the clear-cut identification of the power-law behavior at high frequencies, well within the investigated and accessible spectral range. Such a firm experimental finding allows an unprecedented discussion of the pressure dependence of electronic correlation effects. Our data can be consistently explained within the framework of the TLL theory and share several common properties with other prototype 1D systems, like the linear-chain organic materials.

\begin{acknowledgments}
The authors wish to thank T. Giamarchi for fruitful
discussions. One of us (A.S.) wishes to acknowledge the
scholarship of the Della Riccia Foundation. This work has been
supported by the Swiss National Foundation for the Scientific
Research as well as by the NCCR MaNEP pool and also by the Department of
Energy, Office of Basic Energy Sciences under contract
DE-AC02-76SF00515.
\end{acknowledgments}


\begin{thebibliography}{99}
\bibitem{comment1} Our previous investigations, reported in Ref. \onlinecite{Sacchettiprl}, were extending down to 2700 cm$^{-1}$ but only up to 8000 cm$^{-1}$, while the present work focuses the attention on the high frequency interval up to 1.5x10$^4$ cm$^{-1}$.
\bibitem{comment2} While the power-law behavior is reproducible in several data set, we emphasize nevertheless that $\sigma_1(\omega)\sim\omega^{-\eta}$ is established over a limited energy interval (i.e., less than a decade).
\end{thebibliography}
\end{document}